\documentclass[twoside,fleqn]{article}
\usepackage{espcrc2}
\usepackage{graphicx}
\newcommand{\be}{\begin{equation}}
\newcommand{\ee}{\end{equation}}

\newcommand{\numberofconfigs}{110}

\title{
\vspace{-2.8cm}
\hfill \rm \null \hfill
\hbox{\normalsize ADP-03-145/T584} \\
\hfill \hbox{\normalsize DESY-03-207} \\
\vspace{1.65cm}
 Hybrid Meson Spectrum from the FLIC action }
\author{J.N.~Hedditch\address[CSSM]{Special Research Center for the
    Subatomic Structure of Matter, and		\\
    Department of Physics, University of Adelaide
    Adelaide SA 5005  Australia},
    D.~B.~Leinweber\addressmark[CSSM],
    A.~G.~Williams\addressmark[CSSM]
    and
    J.~M.~Zanotti\addressmark[CSSM]\address[DESY]{John von
    Neumann-Institut f\"ur Computing NIC, \\
    Deutsches Elektronen-Synchrotron DESY, D-15738 Zeuthen, Germany}
} 
 
\begin{document}

\begin{abstract}
The spectral properties of hybrid meson interpolating fields are
investigated.  The quantum numbers of the meson are carried by
smeared-source fermion operators and highly-improved chromo-electric
and -magnetic field operators composed with APE-smeared links.  The
effective masses of standard and hybrid operators indicate that the
ground state meson is effectively isolated using both standard and
hybrid interpolating fields.  Focus is placed on interpolating fields
in which the large spinor components of the quark and antiquark
fields are merged.  In particular, the effective mass of the exotic
$1^{-+}$ meson is reported. Further, we report some values for excited
mesonic states using a variational process.
\end{abstract}

\maketitle

\section{INTRODUCTION}

Major experimental efforts are currently aimed at determining the
possible existence of exotic mesons; mesons having quantum numbers
that cannot be carried by the minimal Fock space component of a
quark-antiquark pair.  Of particular mention is the proposed program
of the GlueX collaboration associated with the forthcoming upgrade of
the Jefferson Laboratory facility.  The observation of exotic states
and the determination of their properties would elucidate aspects of
QCD which are relatively unexplored.

The quantum numbers $J^{PC} = 0^{+-},\ 0^{--},\ 1^{-+},$ etc.  cannot
be carried by a quark-antiquark pair in a ground-state $S$-wave.
Lattice QCD calculations exploring the non-trivial role of explicit
gluon degrees of freedom in carrying the quantum numbers of the meson
suggest that exotic meson states do indeed exist and have a mass the
order of 2~GeV \cite{importantPapers}.  These findings are further
supported here.  

\section{SIMULATION METHODOLOGY}

Operators carrying exotic quantum numbers can be constructed by
merging standard local interpolating fields $\overline{q}^a(x) \Gamma
q^a(x)$ with chromo-electric, $E_i^{ab}(x)$, or chromo-magnetic
fields, $B_i^{ab}(x)$.  The $J^{PC}$ quantum numbers of the
interpolator are derived from the direct product of those associated
with the quark bilinear and $E_i^{ab}\ (1^{--})$ or $B_i^{ab}\
(1^{+-})$.
For example, combining the vector current of the $\rho$ meson with a
chromo-magnetic field, $1^{--} \otimes 1^{+-}$ provides $0^{-+} \oplus
1^{-+} \oplus 2^{-+}$ with the $0^{-+}$: $\bar q^a \gamma_i q^b
B_i^{ab}$ ($\pi$ meson) and the $1^{-+}$: $\epsilon_{ijk} \bar q^a
\gamma_i q^b B_j^{ab}$ ({\bf exotic}).
We restrict ourselves to the lowest energy-dimension operators,
as these provide better signal with smaller statistical errors.
Table \ref{interpolators} summarizes the standard and hybrid
interpolating fields explored herein.  

\begin{table*}
\caption{$J^{PC}$ quantum numbers and their associated meson
  interpolating fields.}
\label{interpolators}
\begin{tabular}{cccc}
\hline
\noalign{\smallskip}
$0^{++}$ & $0^{+-}$ & $0^{-+}$ & $0^{--}$ \\
\hline
\noalign{\smallskip}
$\bar{q}^a q^a$ & $\bar{q}^a \gamma_4 q^a$ &  $\bar{q}^a \gamma_5 q^a$ & $-i \bar{q}^a \gamma_5 \gamma_j E^{ab}_j q^b$ \\
$- i \bar{q}^a \gamma_j E^{ab}_j q^b$ & $ \bar{q}^a \gamma_5 \gamma_j B^{ab}_j q^b$ & $\bar{q}^a \gamma_5 \gamma_4 q^a$ \\
$-  \bar{q}^a \gamma_j \gamma_4 \gamma_5 B^{ab}_j q^b$ & & $- \bar{q}^a \gamma_j B^{ab}_j q^b$ \\
$-  \bar{q}^a \gamma_j \gamma_4 E^{ab}_j q^b$ & & $- \bar{q}^a \gamma_4 \gamma_j B^{ab}_j q^b$        \\
\noalign{\smallskip}
\hline
\noalign{\medskip}
\hline
\noalign{\smallskip}
$1^{++}$ & $1^{+-}$ & $1^{-+}$ & $1^{--}$ \\
\hline
\noalign{\smallskip}
  $- i \bar{q}^a \gamma_5 \gamma_j q^a$
& $- i \bar{q}^a \gamma_5 \gamma_4 \gamma_j q^a$ 
& $\bar{q}^a \gamma_4 E^{ab}_j q^b$
& $- i \bar{q}^a \gamma_j q^a$\\
  $ i \bar{q}^a \gamma_4 B^{ab}_j q^b$
& $i \bar{q}^a B^{ab}_j q^b $ 
& $- \epsilon_{jkl} \bar{q}^a \gamma_k B^{ab}_l q^b$
& $\bar{q}^a E^{ab}_j q^b$\\
  $ i \epsilon_{jkl} \bar{q}^a \gamma_k E^{ab}_l q^b$
& $\bar{q}^a \gamma_5 E^{ab}_j q^b $
& $\epsilon_{jkl} \bar{q}^a \gamma_4 \gamma_k B^{ab}_l q^b$
& $ - i \bar{q}^a \gamma_5 B^{ab}_j q^b $\\
   $ i \epsilon_{jkl} \bar{q}^a \gamma_k \gamma_4 E^{ab}_l q^b$
& $\bar{q}^a \gamma_5 \gamma_4 E^{ab}_j q^b$
& $- i \epsilon_{jkl} \bar{q}^a \gamma_5 \gamma_4 \gamma_k E^{ab}_l q^b$
& $i \bar{q}^a \gamma_4 \gamma_5 B^{ab}_j q^b$\\
\noalign{\smallskip}
\hline
\end{tabular}
\end{table*}

The formulation of effective interpolating fields for the creation and
annihilation of exotic meson states continues to be an active area of
research.
 
For example, one can generalize the structure of the interpolating
fields further to include nonlocal components where link paths are
incorporated to maintain gauge invariance and carry the nontrivial
quantum numbers of the gluon fields.  In this case, numerous quark
propagators are required for each gauge field configuration rendering
the approach computationally expensive. 
 
Here we consider local interpolating fields.  Gauge-invariant Gaussian
smearing \cite{Gusken:qx,Zanotti:2003fx} is applied at the fermion
source ($t=3$), and local sinks are used to maintain strong signal in
the two-point correlation functions.  Chromo-electric and -magnetic
fields are created from APE-smeared links \cite{ape} at both the
source and sink using the highly-improved ${\mathcal O}(a^4)$-improved
lattice field strength tensor \cite{Bilson-Thompson:2002jk}.  In this
study, the smearing fraction $\alpha = 0.7$ (keeping 0.3 of the
original link) and the process of smearing and $SU(3)$ link projection
is iterated four times \cite{Bonnet:2000dc}.  This amount of smearing
is sufficient to provide a meaningful topological charge and appears
to be suitable for the creation of exotic mesons.  As such, the
results presented here supersede an earlier presentation of hybrid
meson masses \cite{Hedditch:2003dm}.

Propagators are generated using the fat-link irrelevant clover (FLIC)
fermion action \cite{FATJAMES} where the irrelevant Wilson and clover
operators of the fermion action are constructed using fat links while
the relevant operators use the untouched (thin) gauge links.  FLIC
fermions provide a new form of nonperturbative ${\cal O}(a)$
improvement \cite{Leinweber:2002bw,inPrep} where near-continuum
results are obtained at finite lattice spacing.  Access to the light
quark mass regime is enabled by the improved chiral properties of the
lattice fermion action \cite{inPrep}.

Excited states are extracted using a variational technique, corresponding
to a construction of optimal linear combinations of the original operators.
For the sake of completeness, we shall discuss this here, 
in direct analogy to the procedure described in \cite{Melnitchouk:2002eg}.

\section{ANALYTICAL PROCESS}
\label{Analysis}

Consider the momentum-space meson two-point function
for $t > 0$,
\begin{equation}
{G}_{ij}(t,\vec{p}) = \sum_{\vec{x}} e^{-i\vec{p} \cdot \vec{x}}
\langle\Omega|
  \chi_i(t,\vec{x}) {\chi}^{\dagger}_j(0,\vec 0)
|\Omega\rangle\ ,
\end{equation} where $i,j$ label the different interpolating fields
and we focus on Lorentz scalar interpolators for simplicity.
At the hadronic level,
\begin{eqnarray*}
{G}_{ij}(t,\vec{p})\!\! &=&\!\! \sum_{\vec{x}} e^{-i\vec{p} \cdot \vec{x}}
\sum_{H, {p}^{\prime} }
\langle\Omega| \chi_i(t,\vec{x})|H, p' \rangle \nonumber \\
\!\!&\times&\!\!
\langle H, p' | {\chi}^{\dagger}_j(0,\vec 0) |\Omega\rangle\ ,
\end{eqnarray*}
where the $|H, p' \rangle$ are a complete set of hadronic states.
\begin{equation}
\sum_{H, {p}^{\prime}}
 |H, p'\rangle\langle H, p'|=I\ .
\end{equation}
We can make use of translational invariance to write this as
\begin{eqnarray}
&&\sum_{\vec{x}, H, {p}^{\prime}} e^{-i\vec{p} \cdot \vec{x}}
\left\langle\Omega\left|
  \chi_i(0)
  e^{i\hat{\vec{P}} \cdot \vec{x}} e^{-\hat{H}t}
\right|H, p' \right\rangle \times \nonumber \\
&&\quad \left\langle H, p' \left| {\chi}^{\dagger}_j(0)
\right| \Omega \right\rangle                            \nonumber \\
&& = \sum_{H} e^{-{E_{H}t}}
\left\langle\Omega| \chi_i |H, p \rangle
 \langle H, p | {\chi}^{\dagger}_j|\Omega\right\rangle\ .
\end{eqnarray}

It is convenient in the following discussion to label the states
which have the $\chi$ interpolating field quantum numbers 
as $|H_{\alpha}\rangle$ for
$\alpha=1,2,\cdots,N$.  In general the number of states,
$N$, in this tower of excited states may be infinite, but we
will only ever need to consider a finite set of the lowest such
states here.
After selecting zero momentum, $\vec p=0$,
\begin{equation}
\label{eqn:Gijequation}
G_{ij}(t)
 \equiv {G}_{ij}(t,\vec 0)
 = \sum_{\alpha=1}^{N} e^{-{m_{\alpha}}t}
   \lambda^{\alpha}_i {\lambda}^{\dagger\alpha}_j\ ,
\end{equation}
where $\lambda^{\alpha}_i$ and ${\lambda}^{\dagger\alpha}_j$ are
coefficients denoting the
couplings of the interpolating fields $\chi_i$ and ${\chi}^{\dagger}_j$,
respectively, to the state $\left|H_{\alpha}\right\rangle$.
If we use identical source and sink interpolating fields then it
follows from the definition of the coupling strength that
${\lambda}^{\dagger\alpha}_j = (\lambda^{\alpha}_j)^*$ and from
Eq.~(\ref{eqn:Gijequation}) we see that $G_{ij}(t)=[G_{ji}(t)]^*$,
i.e., $G$ is a Hermitian matrix.  If, in addition, we use only real
coefficients in the link products, then $G$ is a real symmetric
matrix.  For the correlation matrices that we construct we have real
link coefficients but we use smeared sources and point sinks and so in
our calculations $G$ is a real but non-symmetric matrix.  Since $G$ is
a real matrix for the infinite number of possible choices of
interpolating fields with real coefficients, then we can take
$\lambda^{\alpha}_i$ and ${\lambda}^{\dagger\alpha}_j$ to be real
coefficients here without loss of generality.  In constructing
correlation functions, we effectively average over $\{ U \}$ and 
$\{ U^* \}$ configurations to ensure $\lambda^{\alpha}_i$ is purely
real, even on a finite ensemble of gauge field configurations
\cite{Melnitchouk:2002eg}.

Now, let us  consider the ideal case where we have $N$
interpolating fields with the same quantum numbers, but which
give rise to $N$ linearly independent states when acting on the
vacuum.  In this case we can construct $N$ ideal interpolating
source and sink fields which perfectly isolate the $N$ individual
hadron states $|H_\alpha\rangle$, i.e.,
\begin{eqnarray}
{\phi}^{\dagger\alpha} &=& \sum_{i=1}^N 
          u^{\alpha}_i\ {\chi}^{\dagger}_i\ , \\
\phi^{\alpha} &=& \sum_{i=1}^N v^{\ast \alpha}_i\ \chi_i\ ,
\end{eqnarray}
\label{lincomIF}
such that
\begin{eqnarray}
\left\langle H_{\beta}\right| {\phi}^{\dagger\alpha}
\left| \Omega\right\rangle
&=& \delta_{\alpha\beta}\ {z}^{\dagger\alpha}\ , 
\label{PhiExpressionA} \\
\left\langle \Omega \right | \phi^{\alpha}
\left| H_{\beta}\right\rangle
&=& \delta_{\alpha\beta}\ z^{\alpha}\ ,
\label{PhiExpressionB}
\end{eqnarray}
where $z^\alpha$ and ${z}^{\dagger\alpha}$ are the coupling strengths
of $\phi^\alpha$ and ${\phi}^{\dagger\alpha}$ to the state
$|H_\alpha\rangle$.
The coefficients $u_i^\alpha$ and $v_i^{\ast \alpha}$ in
Eqs.~(\ref{lincomIF}) may differ when the source and sink have different
smearing prescriptions, again indicated by the differentiation between
$z^\alpha$ and ${z}^{\dagger\alpha}$ (recall $z$ is real).  

For notational convenience for the remainder of this discussion
repeated indices $i,j,k$ are to be understood as being summed over,
whereas $\alpha$ denoting a particular state is not.
At $\vec{p}=0$, it follows that,
\begin{eqnarray}
\label{ActingLeft}
G_{ij}(t)\ u^{\alpha}_j
&=& \left(\sum_{\vec{x}} 
  \left \langle \Omega \right |
  \chi_i {\chi}^{\dagger}_j
  \left| \Omega \right\rangle
\right) u^{\alpha}_j                    \nonumber\\
&=& \lambda^{\alpha}_i {z}^{\dagger\alpha} 
    e^{-m_{\alpha} t} .
\end{eqnarray}
The $t$-dependence in this expression is purely in the exponential
term, leading to the recurrence relationship
\begin{equation}
G_{ij} (t) \, u^{\alpha}_j
= e^{m_{\alpha}} G_{ik} (t+1) \, u^{\alpha}_k\ ,
\label{eveqn}
\end{equation}
which can be rewritten as
\begin{equation}
[G(t+1)]_{ki}^{-1} G_{ij} (t) \, u^{\alpha}_j
= e^{m_{\alpha}}\, u^{\alpha}_k\ .
\label{eveqn2}
\end{equation}
This is the generalized eigenvalue equation for
$[G (t+1)]^{-1} G(t)$ with 
eigenvalues $e^{m_{\alpha}}$ and eigenvectors $u^\alpha$.
Hence the natural logarithms of the eigenvalues of
$[G (t+1)]^{-1} G(t)$ are the masses of the
$N$ hadrons in the tower of excited states for the given quantum numbers
.  The eigenvectors are the coefficients of the
$\chi$ fields providing the optimal linear combination for that
state.  

One can also construct the equivalent left-eigenvalue equation to recover
the $v$ vectors, providing the optimal linear combination of
annihilation interpolators,
\begin{equation}
v^{\ast \alpha}_k G_{kj} (t)
= e^{m_{\alpha}} v^{\ast \alpha}_i G_{ij} (t+1)\ .
\end{equation}
Recalling Eq.~(\ref{ActingLeft}), one finds:
\begin{eqnarray}
G_{ij} (t)\ u^{\alpha}_j
&=& {z}^{\dagger\alpha} \lambda^{\alpha}_i
    e^{-m_{\alpha} t}\ ,                                   \\
v^{\ast \alpha}_i\ G_{ij} (t)
&=& {z}^{\alpha} {\lambda}^{\dagger\alpha}_j e^{-m_{\alpha} t }\ , \\
v^{\ast \alpha}_k\ G_{kj} (t) G_{il} (t)\ u^{\alpha}_l
&=& z^{\alpha} {z}^{\dagger\alpha} \lambda^{\alpha}_i 
{\lambda}^{\dagger\alpha}_j e^{-2m_{\alpha} t}\ .
\label{preprojection}
\end{eqnarray}
The definitions of Eqs.~(\ref{PhiExpressionA}) and
(\ref{PhiExpressionB}) imply
\begin{equation}
\label{twiddleG}
v^{\ast \alpha}_i\ G_{ij}(t)\ u^{\alpha}_j =
z^{\alpha}{z}^{\dagger\alpha} e^{-m_{\alpha} t } ,
\label{projection}
\end{equation}
indicating the eigenvectors may be used to construct a correlation
function in which a single state mass $m_\alpha$is isolated
and which can be analyzed
using the methods of Section~II. We refer to this as the projected
correlation function in the following.
Combining Eqs.~(\ref{preprojection}) and (\ref{projection}) leads us to
the result
\begin{equation}
\frac{v^{\ast \alpha}_k \ G_{kj}(t) G_{il}(t)\ u^{\alpha}_l}
{v^{\ast \alpha}_k G_{kl}(t) u^{\alpha}_l }
= \lambda^{\alpha}_{i}{\lambda}^{\dagger\alpha}_{j}
  e^{-m_{\alpha} t} 
\ .
\label{eqn:ratios}
\end{equation}
By extracting all $N^2$ such ratios, we can exactly
recover all of the real couplings $\lambda^{\alpha}_{i}$ and 
${\lambda}^{\dagger\alpha}_{j}$ of 
$\chi_i$ and ${\chi}^{\dagger}_j$ respectively to
the state $|H_\alpha\rangle$.

Note that throughout this section no assumptions have been made about
the symmetry properties of $G_{ij}$. This is essential due to our
use of smeared sources and point sinks.

In practice we will only have a relatively small number, $M<N$,
of interpolating fields in any given analysis.  These $M$
interpolators should be chosen to have good overlap with the
lowest $M$ excited states in the tower and we should attempt
to study the ratios in Eq.~(\ref{eqn:ratios}) at early to
intermediate Euclidean times, where the contribution of the
$(N-M)$ higher mass states will be suppressed but where
there is still sufficient signal to allow the lowest $M$ states
to be seen.  This procedure will lead
to an estimate for the masses of each of the lowest $M$ states
in the tower of excited states.  Of these $M$
predicted masses, the highest will in general have the largest
systematic error while the lower masses will be the
most reliably determined.  Repeating the analysis with varying
$M$ and different combinations of interpolating fields
will give an objective measure of the reliability of the
extraction of these masses.

In our case of a modest $2 \times 2$ correlation matrix ($M=2$)
we take a cautious approach to the selection of the
eigenvalue analysis time.
As already explained, we perform the eigenvalue analysis at an early
to moderate Euclidean time where statistical noise is suppressed
and yet contributions from at least the lowest two mass states
is still present.  One must exercise caution in performing the
analysis at too early a time, as more than the desired
$M=2$ states may be contributing to the $2\times 2$ matrix of
correlation functions.

\section{RESULTS}
\label{results}

The following results are based on \numberofconfigs\ mean-field ${\cal
O}(a^2)$-improved Luscher-Weisz \cite{Luscher:1984xn} gauge fields on
a $16^3 \times 32$ lattice at $\beta = 4.60$ providing a lattice
spacing of $a = 0.122(2)$ fm set by the string tension $\sqrt{\sigma}
= 440$ MeV.

\begin{figure}[t]
\begin{center}
{\includegraphics[height=\hsize,angle=90]{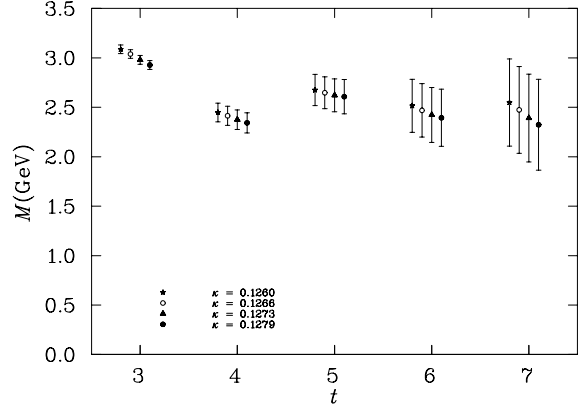}}
\vspace*{-1.4cm}
\caption{Effective mass plot of the $1^{-+}$ exotic meson obtained
from the hybrid interpolating field $\epsilon_{jkl} \bar{q}^a \gamma_k
B^{ab}_l q^b$.  }
\label{1mp22mass}
\end{center}
\vspace*{-1.0cm}
\end{figure}

Of the hybrid interpolators listed in Table \ref{interpolators}, only
the interpolating fields merging the large spinor components of the
quark and antiquark fields provide a clear mass plateau.  The
effective mass plot for the exotic $1^{-+}$ meson is illustrated in
Fig.\ \ref{1mp22mass}, where a plateau at early times is observed
confirming the existence of the exotic $1^{-+}$.

\begin{figure}[t]
\begin{center}
{\includegraphics[height=\hsize,angle=90]{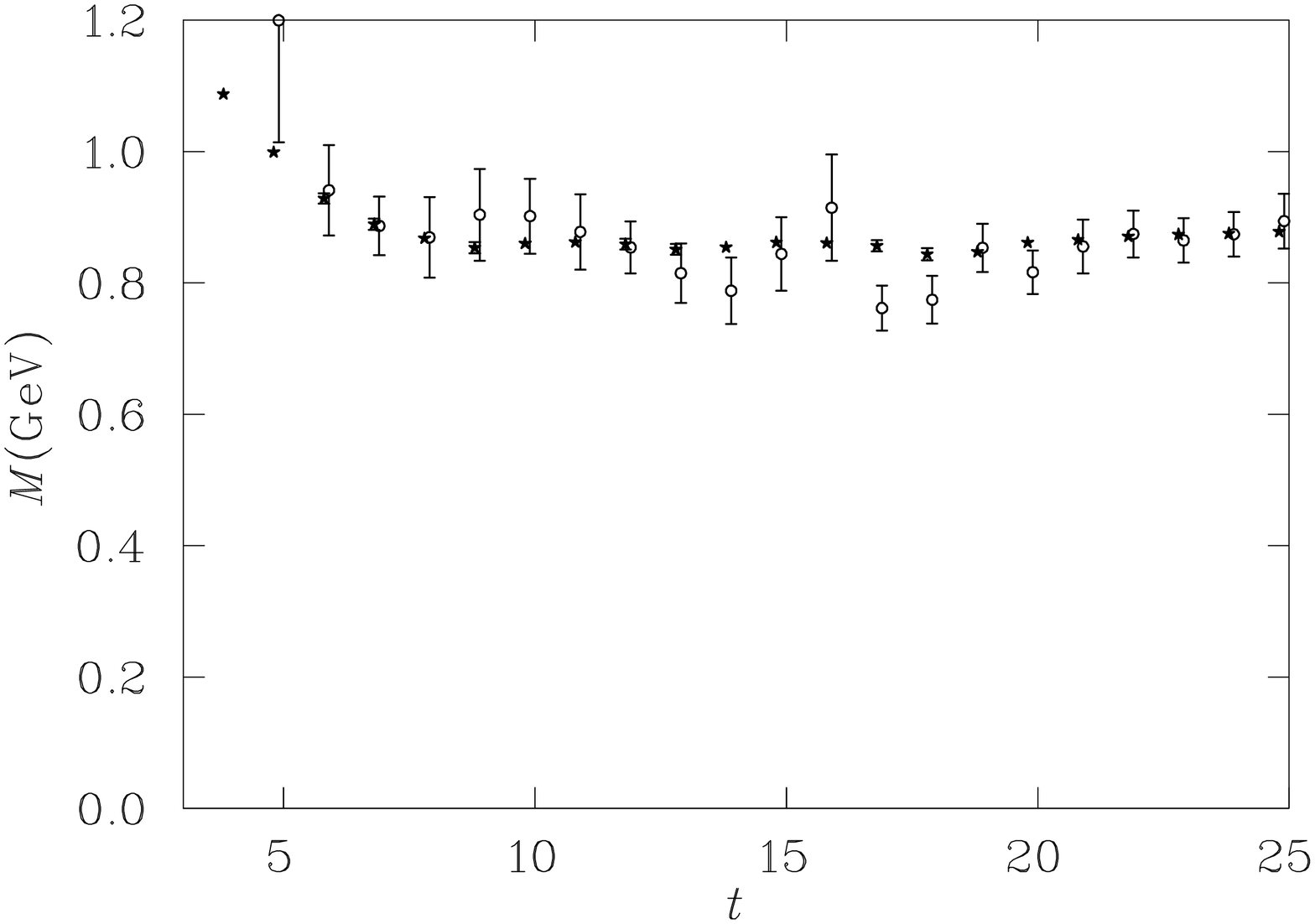}}
\vspace*{-1.4cm}
\caption{Effective mass plot for correlation functions of the standard
pion interpolator $\bar{q}^a \gamma_5 q^a$ and the hybrid pion
interpolator $\bar{q}^a \gamma_j B^{ab}_j q^b$.}
\label{pion11pion33}
\end{center}
\vspace*{-1.0cm}
\end{figure}

\begin{figure}[t]
\begin{center}
{\includegraphics[height=\hsize,angle=90]{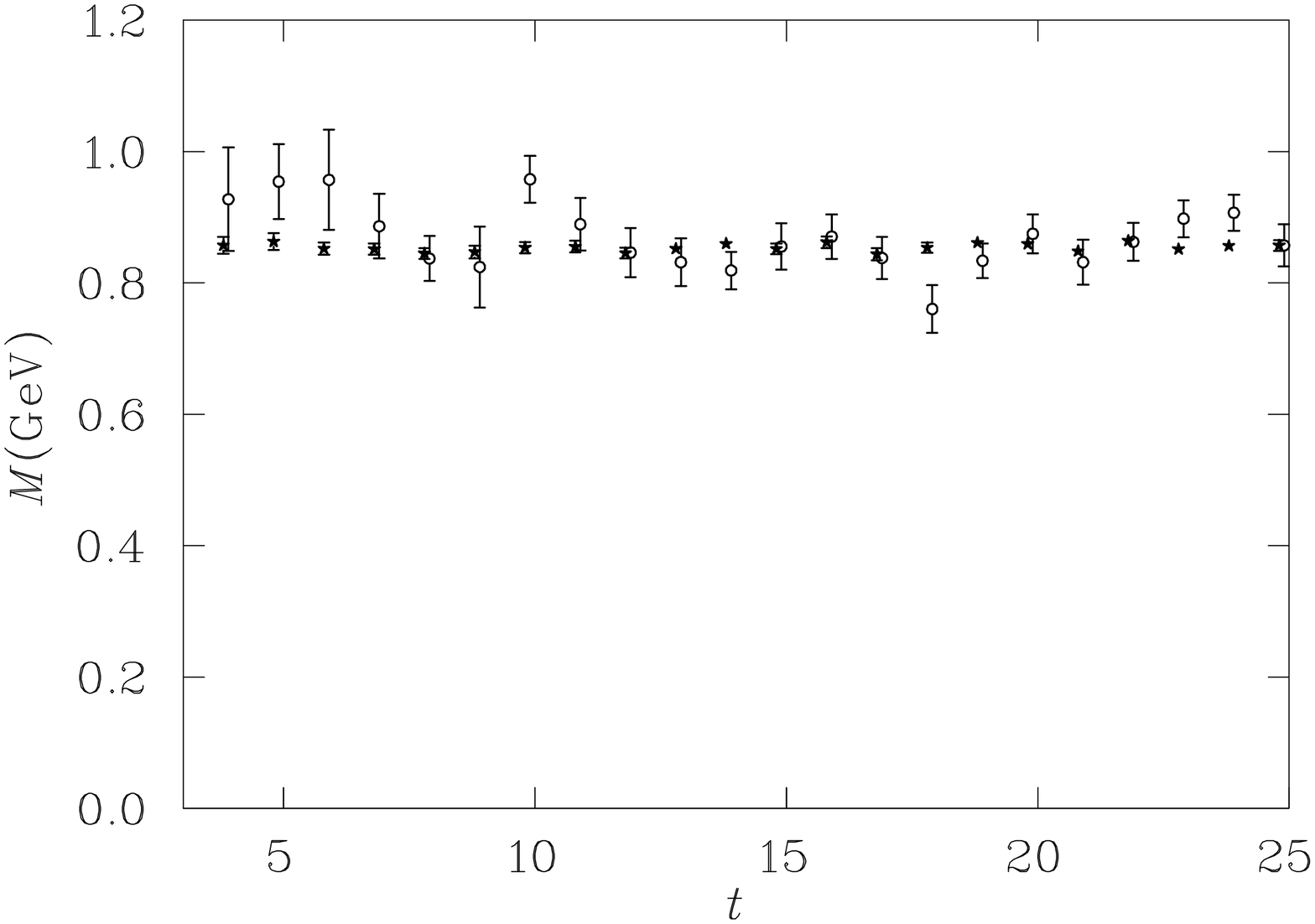}}
\vspace*{-1.4cm}
\caption{Effective mass plot for correlation functions of the standard
axial-vector pion interpolator $\bar{q}^a \gamma_5 \gamma_4 q^a$ and
the hybrid pion interpolator $\bar{q}^a \gamma_4 \gamma_j B^{ab}_j
q^b$.}
\label{pion22pion44}
\end{center}
\vspace*{-1.0cm}
\end{figure}

\begin{figure}[t]
\begin{center}
{\includegraphics[height=\hsize,angle=90]{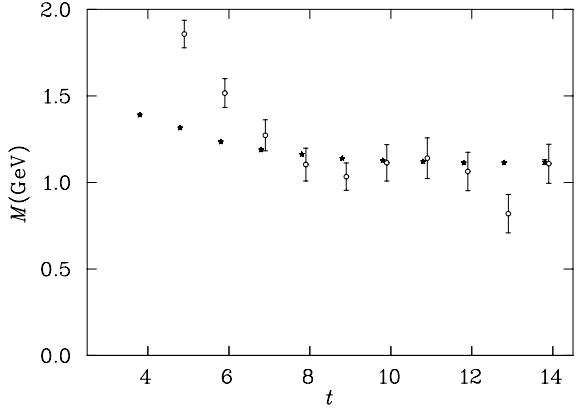}}
\vspace*{-1.4cm}
\caption{Effective mass plot for correlation functions of the
standard$\rho$-meson interpolator $\bar{q}^a \gamma_j q^a $ and the
hybrid $\rho$ interpolator $\bar{q}^a \gamma_4 \gamma_5 B^{ab}_j q^b
$. }
\label{rho11rho33}
\end{center}
\vspace*{-1.0cm}
\end{figure}

Figures \ref{pion11pion33} and \ref{pion22pion44} illustrate the
effective masses $M(t) = - \log (G(t+1)/G(t))$, obtained from the
first and third, and second and fourth, pion ($0^{-+}$) interpolators
of Table \ref{interpolators} for our intermediate quark mass ($m_\pi^2
\sim 0.6\ {\rm GeV}^2$).  Excellent agreement is seen between the
standard and hybrid interpolator-based correlation functions.  Similar
results are seen in Fig.\ \ref{rho11rho33} comparing effective masses
obtained from the first and third $\rho$-meson ($1^{--}$)
interpolators of Table \ref{interpolators}.

\begin{table*}[t]
\caption{Meson masses as a function of the hopping parameter $\kappa$.}
\label{masses_conventional}
\begin{tabular}{rcccccc}
\hline
\noalign{\smallskip}
   &  & \multicolumn{5}{c}{Mass(GeV)} \\
$J^{PC}$&Operator&$\kappa=0.1260$&$\kappa=0.1266$ &
$\kappa=0.1273$ & $\kappa=0.1279$ & $\kappa=0.1286$ 
\\ 
\hline 
\noalign{\smallskip}
$\pi:\, 0^{-+}$  & $\bar{q}^a \gamma_5 q^a$ &
$ 0.965\pm .006$ &
$ 0.887\pm .006$ &
$ 0.789\pm .007$ &
$ 0.696\pm .007$ &
$ 0.570\pm .008$  
\\

        & $\bar{q}^a \gamma_5 \gamma_4 q^a$ & 
$ 0.957\pm .005$ & 
$ 0.879\pm .005$ &
$ 0.781\pm .005$ &
$ 0.688\pm .006$ &
$ 0.563\pm .006$

\\

        & $- \bar{q}^a \gamma_j B^{ab}_j q^b$ & 
$ 0.998\pm .030$ &
$ 0.916\pm .031$ &
$ 0.813\pm .031$ &
$ 0.717\pm .032$ &
$ 0.589\pm .035$ 
\\

       & $- \bar{q}^a \gamma_4 \gamma_j B^{ab}_j q^b$ & 
$ 0.984\pm .033$ &
$ 0.902\pm .034$ &
$ 0.800\pm .035$ &
$ 0.704\pm .037$ &
$ 0.575\pm .039$ 
\\

\noalign{\smallskip}
$b_1:\, 1^{+-}$  &  $- i \bar{q}^a \gamma_5 \gamma_4 \gamma_j q^a$ &
$ 1.713\pm .018$ &
$ 1.671\pm .019$ &
$ 1.623\pm .021$ &
$ 1.583\pm .022$ &
$ 1.541\pm .026$ 
\\

        &  $i \bar{q}^a B^{ab}_j q^b $&
$ 1.685\pm .205$ &
$ 1.621\pm .226$ &
$ 1.525\pm .275$ &
$ 1.398\pm .353$ &
$ 1.120\pm .500$ 
\\
\noalign{\smallskip}
$\rho:\, 1^{--}$  &  $-i \bar{q}^a \gamma_j q^a $&
$ 1.212\pm .009$ &
$ 1.157\pm .010$ &
$ 1.093\pm .012$ &
$ 1.037\pm .014$ &
$ 0.973\pm .019$ 
\\

	& $\bar{q}^a E^{ab}_j q^b $&
$ 1.198\pm .056$ &
$ 1.151\pm .063$ &
$ 1.099\pm .076$ &
$ 1.048\pm .091$ &
$ 0.958\pm .110$ 
\\

        &  $-i \bar{q}^a \gamma_5 B^{ab}_j q^b $&
$ 1.214\pm .050$ &
$ 1.156\pm .053$ &
$ 1.085\pm .058$ &
$ 1.018\pm .064$ &
$ 0.922\pm .077$ 
\\

       &  $i \bar{q}^a \gamma_4 \gamma_5 B^{ab}_j q^b $&
$ 1.125\pm .049$ &
$ 1.060\pm .054$ &
$ 0.982\pm .061$ &
$ 0.907\pm .069$ &
$ 0.799\pm .087$ 
\\
\noalign{\smallskip}
$1^{-+}$ &$- \epsilon_{jkl} \bar{q}^a \gamma_k B^{ab}_l q^b$ &
$ 2.519\pm .186$ &
$ 2.483\pm .192$ &
$ 2.450\pm .200$ &
$ 2.434\pm .207$ &
$ 2.430\pm .211$ 
\\
\noalign{\smallskip}
\hline
\end{tabular}
\end{table*}

Table \ref{masses_conventional} summarizes these preliminary results,
and table \ref{masses_variational} presents a further preliminary
calculation of a pion excited state mass, using the first two pion
operators listed above.  

Further work on this topic will focus on increasing the statistics,
and increasing the number of operators used in the variational
process.

\begin{table*}[t]
\caption{Pion excited state from variational analysis.}
\label{masses_variational}
\begin{tabular}{rccccc}
\hline
    \multicolumn{5}{c}{Mass(GeV)} \\
& $\kappa=0.1260$&$\kappa=0.1266$ &
$\kappa=0.1273$ & $\kappa=0.1279$ & $\kappa=0.1286$ 
\\ 
\hline 
\noalign{\smallskip}
$\pi(1):\,$ &
$0.956\pm .007$ &
$0.878\pm .007$ &
$0.779\pm .008$ &
$0.684\pm .009$ &
$0.555\pm .014$ 
\\
$\pi(2):\, $&
$1.889\pm .064$ &
$1.819\pm .070$ &
$1.730\pm .080$ &
$1.647\pm .096$ &
$1.548\pm .141$ 
\\
\hline

\end{tabular}
\end{table*}

\section{ACKNOWLEDGMENTS}

This research is supported by the Australian National Computing
Facility for Lattice Gauge Theory and the Australian Research Council.


\end{document}